\documentclass[11pt, a4paper,superscriptaddress,aps]{article}
 \usepackage{jheppub}

\usepackage[utf8]{inputenc}

\usepackage{amsfonts}
\usepackage{amssymb}
\usepackage{amsmath}
\usepackage{amsthm}

\usepackage{wrapfig}

\usepackage{bm}
\usepackage{cancel}
\usepackage{bbold}
\usepackage{slashed}
\usepackage{graphicx}
\usepackage{color}
\usepackage{hyperref}
\usepackage{algorithm}
\usepackage{algpseudocode}
\usepackage{tikz}
\usetikzlibrary{shapes}

\usepackage{numprint}
\npdecimalsign{.}
\nprounddigits{2}

\usepackage{xfrac}

\theoremstyle{plain}

\begin{document}

% ==============================================================================

%\title{\Large{Learning QCD: RNNs and the Parton shower}}
\title{\Large{QCD-Aware Recursive Neural Networks for Jet Physics}}
\vspace{1cm}
%\author{\small{\bf Gilles Louppe}\thanks{\texttt{g.louppe@nyu.edu}}}
%\affiliation{New York University}
%\author{\small{\bf Kyunghyun Cho}\thanks{\texttt{kyunghyun.cho@nyu.edu}}}
%\affiliation{New York University}
%\author{\small{\bf Cyril Becot}\thanks{\texttt{kyunghyun.cho@nyu.edu}}}
%\affiliation{New York University}
%\author{\small{\bf Kyle Cranmer}\thanks{\texttt{kyle.cranmer@nyu.edu}}}
%\affiliation{New York University}

\author[a,b,1]{Gilles Louppe%
\note{Currently at University of Li\`ege}}
\author[b]{Kyunghyun Cho}
\author[a,2]{Cyril Becot%
\note{Currently at DESY}}
\author[a,b]{Kyle Cranmer}

\emailAdd{g.louppe@uliege.be}
\emailAdd{kyunghyun.cho@nyu.edu}
\emailAdd{cyril.becot@cern.ch}
\emailAdd{kyle.cranmer@nyu.edu}

\affiliation[a]{New York University, Center for Cosmology \& Particle Physics, 726 Broadway, New York, NY }
\affiliation[b]{New York University, Center for Data Science, 60 5th Ave., New York, NY }
%\affiliation[b]{University of Li\`ege }

\abstract{
Recent progress in applying machine learning for jet physics has been
built upon an analogy between calorimeters and images.
In this work, we present a novel class of recursive
neural networks built instead upon an analogy between QCD and natural languages.
In the analogy, four-momenta are like words and the clustering history of sequential
recombination jet algorithms is like the parsing of a sentence. Our approach works directly with the
four-momenta of a variable-length set of particles, and the jet-based tree structure varies on an event-by-event basis.
Our experiments highlight the flexibility of our method for building task-specific jet embeddings and show that recursive
architectures are significantly more accurate and data efficient than previous
image-based networks. We extend the analogy from individual jets (sentences) to full events (paragraphs),
and show for the first time an event-level classifier operating on all the stable particles produced
in an LHC event.}

%\keywords{keyword one, keyword two}

 \arxivnumber{1702.00748}

\maketitle

% ==============================================================================

\section{Introduction}

By far the most common structures seen in collisions at the Large Hadron Collider (LHC) are
collimated sprays of energetic hadrons referred to as `jets'. These jets are produced
from the fragmentation and hadronization of quarks and gluons as described by quantum
chromodynamics (QCD). Several goals for the LHC are centered around the treatment
of jets, and there has been an enormous amount of effort from both the theoretical and experimental
communities to develop techniques that are able to cope with the experimental realities while
maintaining precise theoretical properties. In particular, the communities have converged on
sequential recombination jet algorithms, methods to study jet substructure, and grooming
techniques to provide robustness to pileup.

One compelling physics challenge is to search for highly boosted standard model particles decaying
hadronically. For instance, if a hadronically decaying $W$ boson is highly boosted, then its
decay products will merge into a single fat jet with a characteristic substructure.
Unfortunately, there is a large background from jets produced by more mundane QCD processes.
For this reason, several jet `taggers' and variables sensitive to jet substructure have been proposed. Initially, this
work was dominated by techniques inspired by our intuition and knowledge of QCD; however, more
recently there has been a wave of approaches that eschew this expert knowledge in favor of
machine learning techniques. In this paper, we present a hybrid approach that leverages the structure
of sequential recombination jet algorithms and deep neural networks.

Recent progress in applying machine learning techniques for jet physics has been
built upon an analogy between calorimeters and
images~\cite{Cogan:2014oua,deOliveira:2015xxd,Almeida:2015jua,Baldi:2016fql,Guest:2016iqz,Barnard:2016qma,Komiske:2016rsd,Kasieczka:2017nvn}.
These methods take a variable-length set of 4-momenta and project them into a
fixed grid of $\eta-\phi$ towers or `pixels' to produce a `jet image'.
The original jet classification problem, hence, reduces to an image
classification problem, lending itself to deep
convolutional networks and other machine learning algorithms.
Despite their promising results, these models suffer
from the fact that they have many free parameters and that they require large amounts of
data for training.
More importantly, the projection of jets into images also loses information, which
impacts classification performance.
The most obvious way to address this issue is
to use a recurrent neural network to process a sequence of 4-momenta as they are. However,
it is not clear how to order this sequence. While $p_T$ ordering is common in
many contexts~\cite{Guest:2016iqz}, it does not capture important angular information critical for
understanding the subtle structure of jets.

In this work, we propose instead a solution for jet classification based on an
analogy between QCD and natural languages, as inspired by several works from
natural language
processing~\cite{goller1996learning,socher2011parsing,socher2011semi,cho2014properties,cho2014learning,chen2015sentence}.
Much like a sentence is composed of words following a syntactic
structure organized as a parse tree, a jet is also composed of
4-momenta following a structure dictated by QCD and organized via the clustering history of a
sequential recombination jet algorithm. More specifically, our approach uses `recursive'
networks  where the topology of the network is given by the  clustering history of a
sequential recombination jet algorithm, which varies on an event-by-event basis.
This event-by-event adaptive structure can be contrasted with the `recurrent'
networks that operate purely on sequences (see e.g., \cite{dlbook}). The network is
therefore given the 4-momenta without any loss of information, in a way that
also captures substructures, as motivated by physical theory.

It is convenient to think of the recursive neural network as providing a `jet embedding',
which maps a set of 4-momenta into  $\mathbb{R}^q$. This embedding has fixed length and can be
fed into a subsequent network used for classification or regression. Thus the procedure can be used for
jet tagging or estimating parameters that characterize the jet, such as the masses of resonances buried inside the jet.
Importantly, the embedding and the subsequent network can be trained jointly so that the embedding is optimized for the
task at hand.

Extending the natural language analogy paragraphs of text are sequence of sentences, just
as event are sequence of jets. In particular, we propose
to embed the full particle content of an event by feeding a sequence of jet-embeddings into a recurrent network.
As before, this event-level embedding can
be fed into a subsequent network used for classification or regression.
To our knowledge, this represents the first machine learning model operating on all the detectable particles in an event.

The remainder of the paper is structured as follows. In Sec.~\ref{S:problem}, we
formalize the classification tasks at the jet-level and event-level.
We describe the proposed recursive network architectures in Sec.~\ref{sec:re}
and detail the data samples and preprocessing used in our experiments in
Sec.~\ref{sec:data}. Our results are summarized and discussed first in Sec.~\ref{S:JetResults}
for experiments on a jet-level classification problem,
and then in Sec.~\ref{S:EventResults} for experiments on an event-level
classification problem. In
Sec.~\ref{sec:related}, we relate our work to close contributions from deep
learning, natural language processing, and jet physics. Finally, we gather our
conclusions and directions for further works in Sec.~\ref{sec:conclusions}.

% ==============================================================================

\section{Problem statement}\label{S:problem}

%In this section, we formalize supervised learning tasks at the jet-level and event-level.
We describe a collision event $\mathbf{e}\in \mathcal{E}$ as being composed of a varying
number of particles, indexed by $i$, and where each particle is
represented by its 4-momentum vector $\mathbf{v}_i \in \mathbb{R}^4$, such that
$\mathbf{e} = \{ \mathbf{v}_i | i = 1, \dots, N \}$.

The 4-momenta in each event can be clustered into jets with a sequential recombination jet algorithm
that recursively combines (by simply adding their 4-momenta) the pair $i, i'$ that minimize
\begin{equation}
d^\alpha_{ii'} = \min(p_{ti}^{2\alpha}, p_{ti'}^{2\alpha}) \frac{\Delta R^2_{ii'}}{R^2}
\end{equation}
%(where $\Delta R^2_{ii'}=\Delta \eta_{ii'}^2 +\Delta\phi_{ii'}^2$)
while $d^\alpha_{ii'}$ is less than $\min(p_{ti}^{2\alpha}, p_{ti'}^{2 \alpha})$~\cite{Cacciari:2008gp,Salam:2009jx}.
These sequential recombination algorithms have three hyper-parameters: $R$, $p_{t,\textrm{min}}$, $\alpha$,
and jets with $p_t < p_{t,\textrm{min}}$ are discarded.
At that point, the jet
algorithm has clustered $\mathbf{e}$ into $M$ jets, each of which can be
represented by a binary tree $\mathbf{t}_j \in \mathcal{T}$ indexed by  $j=1,
\dots, M$ with $N_{j}$  leaves (corresponding to a subset of the
$\mathbf{v}_i$).
In the following, we will consider the
specific cases where $\alpha=1,0,-1$, which respectively correspond to
the $k_t$, Cambridge-Aachen and anti-$k_t$ algorithms.

In addition to jet algorithms, we consider a `random' baseline that
corresponds to recombining particles at random to form random binary  trees $\mathbf{t}_j$, along
with  `asc-$p_T$' and `desc-$p_T$' baselines, which correspond to degenerate binary trees formed from the sequences
of particles sorted respectively in ascending and descending order of $p_T$.

For jet-level classification or regression, each jet $\mathbf{t}_j \in {\cal T}$ in the
training data comes with labels or regression values $y_j \in {\cal
Y}^\text{jet}$. In this framework, our goal is to build a predictive model
$f^\text{jet} : {\cal T} \mapsto  {\cal Y}^\text{jet}$ minimizing some loss
function ${\cal L}^\text{jet}$. Similarly, for event-level classification or
regression, we assume that each collision event $\mathbf{e}_l \in {\cal E}$ in the training
data comes with labels or regression values $y_l \in {\cal Y}^\text{event}$,
and our goal is to build a predictive model $f^\text{event} : {\cal
E} \mapsto  {\cal Y}^\text{event}$ minimizing some loss function ${\cal
L}^\text{event}$.

%% ==============================================================================

\section{Recursive embedding}
\label{sec:re}

\subsection{Individual jets}
\label{sec:re-jets}

Let us first consider the case of an individual jet  whose
particles are topologically structured  as a binary tree $\mathbf{t}_j$, e.g.,
based on a sequential recombination jet clustering algorithm or a
simple sequential sorting in $p_T$.
Let $k = 1, \dots, 2N_j - 1$ indexes the node of the binary tree $\mathbf{t}_j$,
and let the left and right children of node $k$ be denoted by $k_L$ and
$k_R$ respectively. Let also $k_L$ always be
the hardest child of $k$.
By construction, we suppose that leaves $k$ map to particles $i(k)$
while internal nodes correspond to recombinations.
Using these notations, we recursively define
the embedding $\mathbf{h}^\text{jet}_k \in \mathbb{R}^q$ of
node $k$ in $\mathbf{t}_j$ as

\tikzstyle{every node}=[font=\footnotesize]
%\begin{figure}
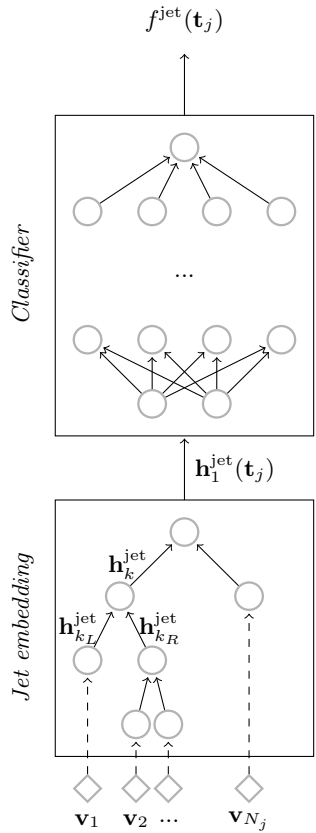
\begin{wrapfigure}{r}{.42\textwidth}
\centering
    \usetikzlibrary{arrows}
    \def\layersep{1}
    \def\stepoffset{4}

    \begin{tikzpicture}[shorten >= 1pt, ->, node distance=\layersep,scale=0.85,rotate=90]
    \tikzstyle{neuron} = [circle, draw=black!30, line width=0.3mm, fill=white];
    \tikzstyle{treenode} = [circle, draw=black!30, line width=0.3mm, fill=white];
    \tikzstyle{particle} = [diamond, draw=black!30, line width=0.3mm, fill=white,scale=0.7];

    %\draw[color=gray, style=dotted] (0,0) grid[xstep=1cm, ystep=1cm] (20cm,6cm);

    % tree 1
    \node[treenode] (tree1-h1) at (8,3) {};
    \node[treenode] (tree1-h2) at (7,4) {};
    \node[treenode] (tree1-h3) at (7,2) {};
    \path[black] (tree1-h2) edge (tree1-h1);
    \path[black] (tree1-h3) edge (tree1-h1);
    \node[treenode] (tree1-h4) at (6, 4.5) {};
    \node[treenode] (tree1-h5) at (6, 3.5) {};
    \path[black] (tree1-h4) edge (tree1-h2);
    \path[black] (tree1-h5) edge (tree1-h2);
    \node[treenode] (tree1-h6) at (5, 3.75)  {};
    \node[treenode] (tree1-h7) at (5, 3.25) {};
    \path[black] (tree1-h6) edge (tree1-h5);
    \path[black] (tree1-h7) edge (tree1-h5);
    \draw (4.5,1) rectangle (8.5, 5);

    \node[particle] (t1-v1) at (4,4.5) {};
    \node at (3.5,4.5) {$\mathbf{v}_1$};
    \path[black,dashed] (t1-v1) edge (tree1-h4);
    \node[particle] (t1-v2) at (4,3.75) {};
    \node at (3.5,3.75) {$\mathbf{v}_2$};
    \path[black,dashed] (t1-v2) edge (tree1-h6);
    \node[particle] (t1-v3) at (4,3.25) {};
    \node at (3.5,3.24) {...};
    \path[black,dashed] (t1-v3) edge (tree1-h7);
    \node[particle] (t1-v4) at (4,2) {};
    \node at (3.5,2) {$\mathbf{v}_{N_j}$};
    \path[black,dashed] (t1-v4) edge (tree1-h3);

    \node[right] at (9,3) {$\mathbf{h}^\text{jet}_1(\mathbf{t}_j)$};
    \path[black] (8.5, 3) edge (9.5,3);

    \node at (7.5,3.875) {$\mathbf{h}^\text{jet}_k$};
    \node at (6.5,4.625) {$\mathbf{h}^\text{jet}_{k_L}$};
    \node at (6.5,3.375) {$\mathbf{h}^\text{jet}_{k_R}$};

    % Classifier f
    \draw (-1+11-0.5,1) rectangle (4+11-0.5,5);
    \foreach \name / \y in {1,...,2}
        \node[neuron] (f-I-\name) at (-1+11,-1-\y+6-0.5) {};

    \foreach \name / \y in {1,...,4}
        \node[neuron] (f-H1-\name) at (-1+\layersep+11,-\y +6-0.5) {};
    \foreach \name / \y in {1,...,4}
        \node[neuron] (f-H2-\name) at (-1+3*\layersep+11,-\y +6-0.5) {};

    \node[neuron] (f-O) at (-1+4*\layersep+11,-3+6) {};

    \foreach \source in {1,...,2}
        \foreach \dest in {1,...,4}
            \path[black] (f-I-\source) edge (f-H1-\dest);

    \foreach \source in {1,...,4}
        \path[black] (f-H2-\source) edge (f-O);

    \node[black] at (1+11,-3+6) {...};
    \path[black] (15-0.5,3) edge (16-0.5,3);
    \node at (16,3) {$f^\text{jet}(\mathbf{t}_j)$};

    \node[rotate=90] at (12,5.5) {\it Classifier};
    \node[rotate=90] at (6.5,5.5) {\it Jet embedding};

    \end{tikzpicture}
    \caption{QCD-motivated recursive jet embedding for classification.
    For each individual jet, the embedding $\mathbf{h}^\text{jet}_1(\mathbf{t}_j)$
    is computed recursively from the root node down to the outer nodes
    of the binary tree $\mathbf{t}_j$.
    The resulting embedding is chained to a subsequent classifier,
    as illustrated in the top part of the figure.
    The topology of the network in the bottom part is distinct
    for each jet and is determined by a sequential recombination jet algorithm (e.g., $k_t$ clustering).
    }
    \label{fig:jet_embedding}
%\end{figure}
\end{wrapfigure}\leavevmode

\begin{align}
\mathbf{h}^\text{jet}_k &=
 \begin{cases}
  \mathbf{u}_k          & \text{if } k \text{ is a leaf} \\
  \sigma \left( W_h
  \begin{bmatrix}
      \mathbf{h}^\text{jet}_{k_L} \\
      \mathbf{h}^\text{jet}_{k_R} \\
      \mathbf{u}_k
  \end{bmatrix} + b_{h} \right)  & \text{otherwise}
 \end{cases} \label{eqn:rec-nn}\\
 \mathbf{u}_k &= \sigma \left( W_u g(\mathbf{o}_k) + b_u \right) \\
 \mathbf{o}_k &=
 \begin{cases}
 \mathbf{v}_{i(k)} & \text{if } k \text{ is a leaf} \\
 \mathbf{o}_{k_L} + \mathbf{o}_{k_R} & \text{otherwise}
 \end{cases}
\end{align}\\

\noindent where
$W_h \in \mathbb{R}^{q \times 3q}$,
$b_h \in \mathbb{R}^q$,
$W_u \in \mathbb{R}^{q \times 4}$
and $b_u \in \mathbb{R}^q$ form together the shared parameters to be learned,
$q$ is the size of the embedding,
$\sigma$ is the ReLU activation function~\cite{nair2010rectified},
and $g$ is a function extracting the kinematic features $p$, $\eta$, $\theta$, $\phi$, $E$, and $p_T$ from the 4-momentum  $\mathbf{o}_k$.

When applying  Eqn.~\ref{eqn:rec-nn} recursively from the root node $k=1$ down to
the outer nodes of the binary tree $\mathbf{t}_j$, the resulting embedding, denoted
$\mathbf{h}^\text{jet}_1(\mathbf{t}_j)$, effectively summarizes the information contained in
the particles forming the jet into a single vector. In particular, this
recursive neural network (RNN) embeds a binary tree of varying shape and size into
a vector of fixed size. As a result, the embedding
$\mathbf{h}^\text{jet}_1(\mathbf{t}_j)$ can now be chained to a
subsequent classifier or regressor to solve our target supervised learning
problem, as illustrated in Figure~\ref{fig:jet_embedding}. All
parameters (i.e., of the recursive jet embedding and of the classifier) are learned jointly using backpropagation through structure
\citep{goller1996learning} to minimize the loss ${\cal L}^\text{jet}$, hence tailoring
the embedding to the specific requirements of the task. Further implementation details,
including an efficient batched computation over distinct binary trees, can be found in Appendix~\ref{sec:impl}.

\clearpage

In addition to the recursive activation of Eqn.~\ref{eqn:rec-nn}, we also
consider and study its extended version equipped  with reset and update
gates (see details in Appendix~\ref{sec:grs}). This gated architecture allows the network to
preferentially pass information along the left-child, right-child, or their combination.

% In addition to the recursive networks over fixed topologies pre-specified by the jet algorithm, we considered a
% few possible generalizations.
% The second generalization we considered was to simultaneously use two topologies
% over the same events, for instance letting information flow along both an
% anti-$k_t$ and $k_t$ network topology ending in a joint embedding
% $(\mathbf{h}^\text{jet}_1^{k_t}, \mathbf{h}^\text{jet}_1^{\mathrm{anti-}k_t}) \in \mathbb{R}^{2q}$.
% Each topology has its own set of parameters describing the recursive activation
% and the resulting joint embedding is fed into a subsequent classifier as before.
% We call this a stereo embedding, and we jointly optimize the classifier and each
% of the two recursive activations.

While we have not performed experiments, we point out that there is an analogous
style of architectures based on  jet algorithms with 2 $\to$ 3
recombinations~\cite{Salam:2009jx,Fischer:2016vfv, Ritzmann:2012ca}.

\subsection{Full events}
\label{sec:re-events}

\tikzstyle{every node}=[font=\normalsize,scale=0.8] %KC added scale
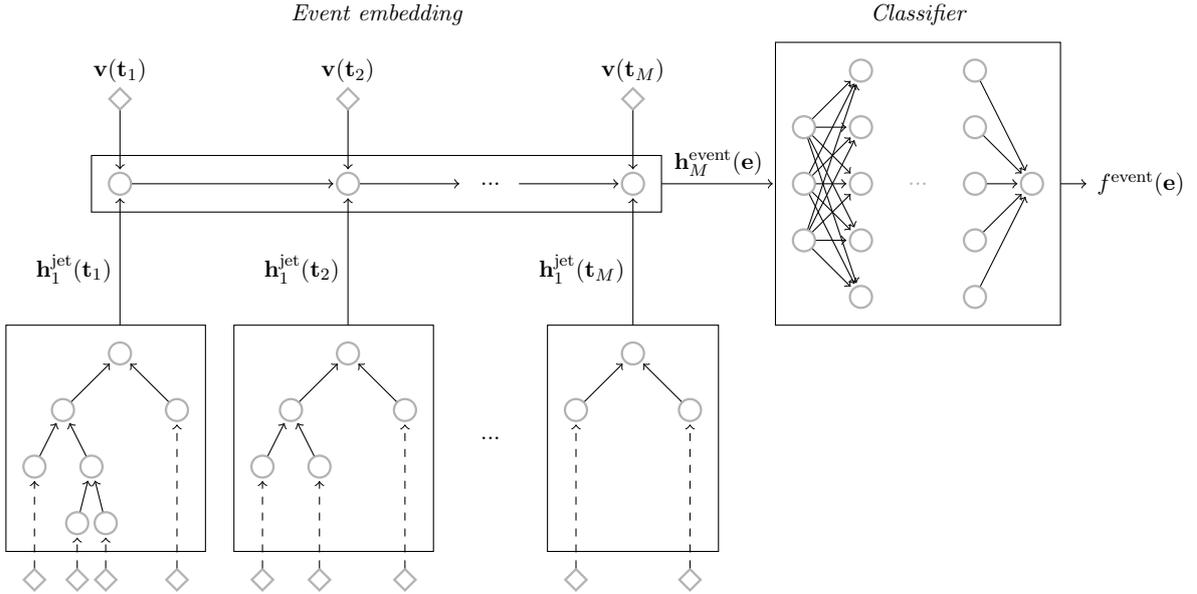
\begin{figure*}
    \usetikzlibrary{arrows}
    \def\layersep{1}
    \def\stepoffset{4}

    \begin{tikzpicture}[shorten >= 1pt, ->, node distance=\layersep,scale=0.75]
    \tikzstyle{neuron} = [circle, draw=black!30, line width=0.3mm, fill=white];
    \tikzstyle{treenode} = [circle, draw=black!30, line width=0.3mm, fill=white];
    \tikzstyle{particle} = [diamond, draw=black!30, line width=0.3mm, fill=white,scale=0.7];

    %\draw[color=gray, style=dotted] (0,0) grid[xstep=1cm, ystep=1cm] (20cm,6cm);

   % %
   %  \node at (6.5,-.5) {\it Jet };
   %  \node at (6.5,-1) {\it embeddings};

    % tree 1
    \node[treenode] (tree1-h1) at (0,0) {};
    \node[treenode] (tree1-h2) at (-1,-1) {};
    \node[treenode] (tree1-h3) at (1,-1) {};
    \path[black] (tree1-h2) edge (tree1-h1);
    \path[black] (tree1-h3) edge (tree1-h1);
    \node[treenode] (tree1-h4) at (-1.5,-2) {};
    \node[treenode] (tree1-h5) at (-0.5,-2) {};
    \path[black] (tree1-h4) edge (tree1-h2);
    \path[black] (tree1-h5) edge (tree1-h2);
    \node[treenode] (tree1-h6) at (-0.75,-3) {};
    \node[treenode] (tree1-h7) at (-0.25,-3) {};
    \path[black] (tree1-h6) edge (tree1-h5);
    \path[black] (tree1-h7) edge (tree1-h5);
    \draw (-2,-3.5) rectangle (1.5, 0.5);

    \node[particle] (t1-v1) at (-1.5,-4) {};
    %\node at (-1.5,-4.5) {$\mathbf{v}_1$};
    \path[black,dashed] (t1-v1) edge (tree1-h4);
    \node[particle] (t1-v2) at (-0.75,-4) {};
    %\node at (-0.75,-4.5) {$\mathbf{v}_2$};
    \path[black,dashed] (t1-v2) edge (tree1-h6);
    \node[particle] (t1-v3) at (-0.25,-4) {};
    %\node at (-0.25,-4.5) {$\mathbf{v}_3$};
    \path[black,dashed] (t1-v3) edge (tree1-h7);
    \node[particle] (t1-v4) at (1,-4) {};
    %\node at (1,-4.5) {$\mathbf{v}_4$};
    \path[black,dashed] (t1-v4) edge (tree1-h3);

    \node[neuron] (gru1) at (0,3) {};
    \path[black] (0,0.5) edge (gru1);
    \node[left] at (0,1.5) {$\mathbf{h}^\text{jet}_1(\mathbf{t}_1)$};
    \node[particle] (jet1-feature) at (0,4.5) {};
    \node at (0,5) {$\mathbf{v}(\mathbf{t}_1)$};
    \path[black] (jet1-feature) edge (gru1);

    % tree 2
    \node[treenode] (tree2-h1) at (0+\stepoffset,0) {};
    \node[treenode] (tree2-h2) at (-1+\stepoffset,-1) {};
    \node[treenode] (tree2-h3) at (1+\stepoffset,-1) {};
    \path[black] (tree2-h2) edge (tree2-h1);
    \path[black] (tree2-h3) edge (tree2-h1);
    \node[treenode] (tree2-h4) at (-1.5+\stepoffset,-2) {};
    \node[treenode] (tree2-h5) at (-0.5+\stepoffset,-2) {};
    \path[black] (tree2-h4) edge (tree2-h2);
    \path[black] (tree2-h5) edge (tree2-h2);
    \draw (2,-3.5) rectangle (5.5, 0.5);

    \node[particle] (t2-v1) at (2.5,-4) {};
    %\node at (2.5,-4.5) {$\mathbf{v}_5$};
    \path[black,dashed] (t2-v1) edge (tree2-h4);
    \node[particle] (t2-v2) at (3.5,-4) {};
    %\node at (3.5,-4.5) {$\mathbf{v}_6$};
    \path[black,dashed] (t2-v2) edge (tree2-h5);
    \node[particle] (t2-v3) at (5,-4) {};
    %\node at (5,-4.5) {$\mathbf{v}_7$};
    \path[black,dashed] (t2-v3) edge (tree2-h3);

    \node[neuron] (gru2) at (0+\stepoffset,3) {};
    \path[black] (\stepoffset,0.5) edge (gru2);
    \node[left] at (\stepoffset,1.5) {$\mathbf{h}^\text{jet}_1(\mathbf{t}_2)$};
    \path[black] (gru1) edge (gru2);
    \node[particle] (jet2-feature) at (0+\stepoffset,4.5) {};
    \node at (0+\stepoffset,5) {$\mathbf{v}(\mathbf{t}_2)$};
    \path[black] (jet2-feature) edge (gru2);

    % ...
    \path[black] (gru2) edge (6,3);
    \node[black] at (6.5,3) {...};
    \node at (6.5,-1.5) {...};

    % tree M
    \node[treenode] (treeM-h1) at (0+3*\stepoffset-3,0) {};
    \node[treenode] (treeM-h2) at (-1+3*\stepoffset-3,-1) {};
    \node[treenode] (treeM-h3) at (1+3*\stepoffset-3,-1) {};
    \path[black] (treeM-h2) edge (treeM-h1);
    \path[black] (treeM-h3) edge (treeM-h1);
    \draw (7.5,-3.5) rectangle (10.5, 0.5);

    \node[particle] (tM-v1) at (8,-4) {};
    %\node at (8,-4.5) {$\mathbf{v}_{N-1}$};
    \path[black,dashed] (tM-v1) edge (treeM-h2);
    \node[particle] (tM-v2) at (10,-4) {};
    %\node at (10,-4.5) {$\mathbf{v}_N$};
    \path[black,dashed] (tM-v2) edge (treeM-h3);

    \node[neuron] (gru3) at (9,3) {};
    \path[black] (7,3) edge (gru3);
    \path[black] (9,0.5) edge (gru3);
    \node[left] at (9,1.5) {$\mathbf{h}^\text{jet}_1(\mathbf{t}_M)$};
    \node[particle] (jet3-feature) at (9,4.5) {};
    \node at (9,5) {$\mathbf{v}(\mathbf{t}_M)$};
    \path[black] (jet3-feature) edge (gru3);

    \draw (-0.5,2.5) rectangle (9.5, 3.5);
    \path[black] (9.5,3) edge (11.5,3);
    \node[above] at (10.5,3) {$\mathbf{h}^\text{event}_M(\mathbf{e})$};
    \node at (4.5,6) {\it Event embedding};

    % Classifier f
    \draw (-1+12.5,-0.5+6) rectangle (4+12.5,-5.5+6);
    \foreach \name / \y in {1,...,3}
        \node[neuron] (f-I-\name) at (-0.5+12.5,-1-\y+6) {};

    \foreach \name / \y in {1,...,5}
        \node[neuron] (f-H1-\name) at (-0.5+\layersep+12.5,-\y +6) {};
    \foreach \name / \y in {1,...,5}
        \node[neuron] (f-H2-\name) at (-0.5+3*\layersep+12.5,-\y +6) {};

    \node[neuron] (f-O) at (-0.5+4*\layersep+12.5,-3+6) {};

    \foreach \source in {1,...,3}
        \foreach \dest in {1,...,5}
            \path[black] (f-I-\source) edge (f-H1-\dest);

    \foreach \source in {1,...,5}
        \path[black] (f-H2-\source) edge (f-O);

    \node[black!30] at (1.5+12.5,-3+6) {...};
    \path[black] (16.5,3) edge (17,3);
    \node[right] at (17,3) {$f^\text{event}(\mathbf{e})$};
    \node at (14,6) {\it Classifier};

    \end{tikzpicture}
    \caption{QCD-motivated event embedding for classification.
    The embedding of an event is computed by feeding the sequence of pairs $(\mathbf{v}(\mathbf{t}_j), \mathbf{h}^\text{jet}_1(\mathbf{t}_j))$ over the jets it is made of,
    where $\mathbf{v}(\mathbf{t}_j)$ is the unprocessed 4-momentum of the jet $\mathbf{t}_j$
    and $\mathbf{h}^\text{jet}_1(\mathbf{t}_j)$ is its embedding. The resulting event-level embedding $\mathbf{h}^\text{event}_M(\mathbf{e})$
    is chained to a subsequent classifier, as illustrated in the right part of the figure.
    %\kyle{Also, would prefer for t1 to be root, similar to jet-level embedding} \gilles{counting backward in the GRU would be strange}
}
    \label{fig:event_embedding}
\end{figure*}

We now embed entire events $\mathbf{e}$ of variable size by
feeding the embeddings of their individual jets to an event-level
sequence-based recurrent neural network.

As an illustrative example, we consider here a gated recurrent
unit~\citep{chung2014empirical} (GRU) operating on the $p_T$ ordered sequence of pairs
$(\mathbf{v}(\mathbf{t}_j), \mathbf{h}^\text{jet}_1(\mathbf{t}_j))$, for $j=1, \dots, M$,
where $\mathbf{v}(\mathbf{t}_j)$ is the unprocessed 4-momentum of the jet $\mathbf{t}_j$
and $\mathbf{h}^\text{jet}_1(\mathbf{t}_j)$ is its embedding.
The final output $\mathbf{h}^\text{event}_M(\mathbf{e})$ (see Appendix~\ref{sec:gru}
for details) of the GRU is chained to a subsequent classifier to solve an
event-level classification task. Again, all parameters (i.e., of the inner jet
embedding function, of the GRU, and of the classifier) are learned jointly using
backpropagation through structure \citep{goller1996learning} to minimize the
loss ${\cal L}^\text{event}$. Figure~\ref{fig:event_embedding} provides a schematic
of the full classification model. In summary, combining two levels of
recurrence provides a QCD-motivated event-level embedding that effectively operates
at the hadron-level for all the particles in the event.

In addition and for the purpose of comparison, we also consider the simpler
baselines where i) only the 4-momenta $\mathbf{v}(\mathbf{t}_j)$
of the jets are given as input to the GRU, without augmentation with
their embeddings, and ii) the 4-momenta $\mathbf{v}_i$ of
the constituents of the event are all directly given as input to the GRU,
without grouping them into jets or providing the jet embeddings.

% ==============================================================================

\section{Data, Preprocessing and Experimental Setup}
\label{sec:data}

In order to focus attention on the impact of the network architectures and the
projection of input 4-momenta into images, we consider the same boosted $W$
tagging example as used in
Refs.~\cite{Cogan:2014oua,deOliveira:2015xxd,Baldi:2016fql,Barnard:2016qma}. The
signal ($y=1$) corresponds to a hadronically decaying $W$ boson with $200 < p_T <
500$ GeV, while the background ($y=0$) corresponds to a QCD jet with the same
range of $p_T$.

We are grateful to the authors of Ref.~\cite{Barnard:2016qma} for sharing the
data used in their studies. We obtained both the full-event records from
their \texttt{PYTHIA} benchmark samples, including both the particle-level data and the
towers from the \texttt{DELPHES} detector simulation. In addition, we obtained
the fully processed jet images of 25$\times$25 pixels, which include the initial
$R=1$ anti-$k_t$ jet clustering and subsequent trimming, translation,
pixelisation, rotation, reflection, cropping, and normalization preprocessing
stages detailed in Ref.~\cite{deOliveira:2015xxd,Barnard:2016qma}.

Our training data was collected by sampling from the original data  a total of
100,000 signal and background jets with equal prior. The testing data
was assembled similarly by sampling 100,000 signal and background jets, without overlap
with the training data.  For direct comparison with Ref.~\cite{Barnard:2016qma},
performance is evaluated at test time within the restricted window of $250 < p_T <
300$ and $50 \leq m \leq 110$, where the signal and background jets are re-weighted to produce flat $p_T$ distributions.
Results are reported in terms of the area under the ROC curve (ROC AUC)
and of background rejection (i.e., 1/FPR) at 50\% signal efficiency ($R_{\epsilon=50\%}$).
Average scores reported include uncertainty estimates
that come from training 30 models with distinct initial random seeds.
About 2\% of the models had technical problems during training (e.g., due to numerical errors), so we applied a
simple algorithm to ensure robustness: we discarded models whose $R_{\epsilon=50\%}$ was
outside of 3 standard deviations of the mean, where the mean and standard
deviation were estimated excluding the five best and worst performing models.

For our jet-level experiments we consider as input to the classifiers the
4-momenta $\mathbf{v}_i$ from both the particle-level data and the \texttt{DELPHES}
towers. We also compare the performance with and without the projection of those
4-momenta into images. While the image data already included the full
pre-processing steps, when considering particle-level and tower inputs we
performed the initial $R=1$ anti-$k_t$ jet clustering to identify the
constituents of the highest $p_T$ jet $\mathbf{t}_1$ of each event, and then
performed the subsequent translation, rotation, and reflection pre-processing
steps (omitting  cropping and normalization). When processing the image data, we
inverted the normalization that enforced the sum  of the squares of the pixel
intensities be equal to one.\footnote{In Ref.~\cite{deOliveira:2015xxd}, the
jet images did not include the \texttt{DELPHES} detector simulation, they were
comparable to our \textit{particle} scenario with the additional discretization into pixels. }

For our event-level experiments we were not able to use the data from
Ref.~\cite{Barnard:2016qma} because the signal sample corresponded to $pp \to
W(\to J) Z(\to \nu\bar{\nu})$ and the background to $pp\to jj$. Thus the signal
was characterized by one high-$p_T$ jet and large missing energy from $Z(\to
\nu\bar{\nu})$ which is trivially separated from the dijet background. For this reason, we
generated our own \texttt{PYTHIA} and \texttt{DELPHES} samples of $pp \to W' \to W(\to J) Z(\to J)$
and QCD background such that both the signal and background have two high-$p_T$
jets. We use $m_{W'}=700$ GeV and restrict $\hat{p}_t$ of the $2\to 2$
scattering process to $300 < \hat{p}_t < 350$ GeV. Our focus is to demonstrate
the scalability of our method to all the particles or towers in an event, and not to
provide a precise statement about physics reach for this signal process. In this
case each event $\mathbf{e}$ was clustered by the same anti-$k_t$
algorithm with $R=1$, and then the constituents of each jet were treated as in Sec.~\ref{sec:re-jets} (i.e., reclustered using $k_t$ or a sequential ordering in $p_T$ to
provide the network topology for a non-gated embedding). Additionally, the constituents
of each jet were pre-processed with translation, rotation, and reflection as in
the individual jet case.  Training was carried out on
a dataset of 100,000 signal and background events with equal prior. Performance
was evaluated on an independent test set of 100,000 other events, as measured
by the ROC AUC and $R_{\epsilon=80\%}$ of the model predictions.
%\gilles{Explain why $R_{\epsilon=80\%}$ instead of $R_{\epsilon=50\%}$}
Again,
average scores are given with uncertainty estimates that come from training
30 models with distinct initial random seeds.

In both jet-level and event-level experiments, the dimension of the embeddings $q$ was set to 40. Training was
conducted using Adam \cite{kingma2014adam} as an optimizer for 25 epochs, with a batch size of $64$
and a learning rate of $0.0005$ decayed by a factor of $0.9$ after every epoch.
These parameters were found to perform best on average, as determined through
an optimization of the hyper-parameters.
Performance was monitored during training on a validation set of $5000$ samples
to allow for early stopping and prevent from overfitting.

%pythia version 8.2.19
%Default PDF (NNPDF2.3 QCD+QED LO alpha_s(M_Z) = 0.130)
%Parton-level cuts :
%	-pTHatMin = 300.
%	-pTHatMax = 350.
%No cuts on the generators' output
%Default tune (Monash 2013)
%Center of mass energy : 13000.
%Background : all HardQCD processes
%Signal : resonance mass = 700 GeV for both W' and Z'

% ==============================================================================

\section{Experiments with Jet-Level Classification}
\label{S:JetResults}

\begin{table}
\caption{Summary of jet classification performance for several
approaches applied either to particle-level inputs or towers from a \texttt{DELPHES} simulation.
% The MaxOut architecture can only be applied to fixed length inputs and requires a
% pre-processing step that projects 4-vectors into images.
}
\begin{center}
\begin{tabular}{|c| c | c | c |}
\hline  Input  &    Architecture & ROC AUC & $R_{\epsilon=50\%}$  \\ \hline \hline

%%%%%%%%%
      \multicolumn{4}{c}{Projected into images } \\ \hline

    towers 	&                      MaxOut 	&  \bf 0.8418 \hspace{3.2em} 	& -- \\
    towers 	&                      $k_t$ 	& 0.8321 $\pm$ 0.0025 	& \bf 12.7 $\pm$ 0.4 \\
    towers 	&               $k_t$ (gated) 	& 0.8277 $\pm$ 0.0028 	& 12.4 $\pm$ 0.3 \\

\hline    \multicolumn{4}{c}{ Without image preprocessing} \\ \hline
    towers 	&                       $\tau_{21}$	& 0.7644 \hspace{3.85em} 	& 6.79 \hspace{2.4em} \\
    towers 	&                       mass + $\tau_{21}$	& 0.8212 \hspace{3.85em} 	& 11.31 \hspace{2.4em} \\ \hline
    towers 	&                       $k_t$ 	& 0.8807 $\pm$ 0.0010 	& 24.1 $\pm$ 0.6 \\
    towers 	&                         C/A 	& 0.8831 $\pm$ 0.0010 	& 24.2 $\pm$ 0.7 \\
    towers 	&                  anti-$k_t$ 	& 0.8737 $\pm$ 0.0017 	& 22.3 $\pm$ 0.8 \\
    towers 	&                   asc-$p_T$ 	& 0.8835 $\pm$ 0.0009 	& \bf 26.2 $\pm$ 0.7 \\
    towers 	&                  desc-$p_T$ 	& \bf 0.8838 $\pm$ 0.0010 	& 25.1 $\pm$ 0.6 \\
    towers 	&                      random 	& 0.8704 $\pm$ 0.0011 	& 20.4 $\pm$ 0.3 \\ \hline
particles 	&                       $k_t$ 	& 0.9185 $\pm$ 0.0006 	& 68.3 $\pm$ 1.8 \\
particles 	&                         C/A 	& \bf 0.9192 $\pm$ 0.0008 	& 68.3 $\pm$ 3.6 \\
particles 	&                  anti-$k_t$ 	& 0.9096 $\pm$ 0.0013 	& 51.7 $\pm$ 3.5 \\
particles 	&                   asc-$p_T$ 	& 0.9130 $\pm$ 0.0031 	& 52.5 $\pm$ 7.3 \\
particles 	&                  desc-$p_T$ 	& 0.9189 $\pm$ 0.0009 	& \bf 70.4 $\pm$ 3.6 \\
particles 	&                      random 	& 0.9121 $\pm$ 0.0008 	& 51.1 $\pm$ 2.0 \\
\hline    \multicolumn{4}{c}{With gating (see Appendix~\ref{sec:grs})} \\ \hline

    towers 	&               $k_t$  	& 0.8822 $\pm$ 0.0006 	& 25.4 $\pm$ 0.4 \\
    towers 	&                C/A 	& 0.8861 $\pm$ 0.0014 	& 26.2 $\pm$ 0.8 \\
    towers 	&          anti-$k_t$  	& 0.8804 $\pm$ 0.0010 	& 24.4 $\pm$ 0.4 \\
    towers 	&           asc-$p_T$  	& 0.8849 $\pm$ 0.0012 	& 27.2 $\pm$ 0.8 \\
    towers 	&          desc-$p_T$  	& \bf 0.8864 $\pm$ 0.0007 	& \bf 27.5 $\pm$ 0.6 \\
    towers 	&             random    & 0.8751 $\pm$ 0.0029 	& 22.8 $\pm$ 1.2 \\ \hline
particles 	&               $k_t$  	& 0.9195 $\pm$ 0.0009 	& 74.3 $\pm$ 2.4 \\
particles 	&                C/A  	& \bf 0.9222 $\pm$ 0.0007 	& 81.8 $\pm$ 3.1 \\
particles 	&          anti-$k_t$  	& 0.9156 $\pm$ 0.0012 	& 68.3 $\pm$ 3.2 \\
particles 	&           asc-$p_T$  	& 0.9137 $\pm$ 0.0046 	& 54.8 $\pm$ 11.7 \\
particles 	&          desc-$p_T$  	& 0.9212 $\pm$ 0.0005 	& \bf 83.3 $\pm$ 3.1 \\
particles 	&             random	& 0.9106 $\pm$ 0.0035 	& 50.7 $\pm$ 6.7 \\

\hline
\end{tabular}
\end{center}
\label{fig:performance}
\end{table}%%

\subsection{Performance studies}

We carried out performance studies where we varied the following
factors: the projection of the 4-momenta into an image, the source of those
4-momenta, the topology of the RNN, and the presence or
absence of gating.

\paragraph{Impact of image projection}
The first factor we studied was whether or not to project the 4-momenta into
an image as in Refs.~\cite{deOliveira:2015xxd,Barnard:2016qma}. The architectures
used in previous studies required a fixed input (image) representation, and
cannot be applied to the variable length set of input 4-momenta. Conversely, we
can apply the RNN architecture to the discretized image 4-momenta.
Table~\ref{fig:performance} shows that the RNN architecture based on a $k_t$
topology performs almost as well as the MaxOut architecture  in
Ref.~\cite{Barnard:2016qma} when applied to the image pre-processed 4-momenta
coming from \texttt{DELPHES} towers. Importantly the RNN architecture is much
more data efficient. While the MaxOut architecture in
Ref.~\cite{Barnard:2016qma} has 975,693 parameters and was trained with 6M
examples, the non-gated RNN architecture has 8,481 parameters and was trained
with 100,000 examples only.

\begin{figure}
\centering
\includegraphics[width=.6\textwidth]{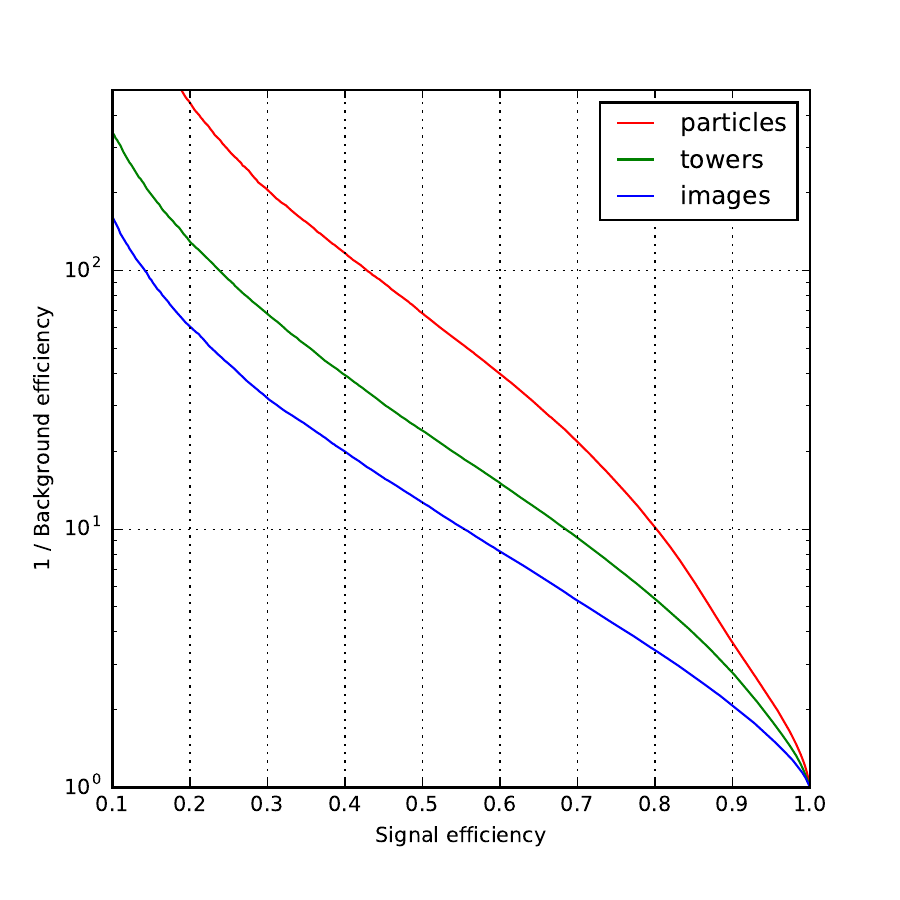}
\caption{Jet classification performance for various input representations of the RNN classifier, using $k_t$ topologies for the embedding.
         The plot shows that there is significant improvement from removing
         the image processing step and that significant gains can be
         made with more accurate measurements of the 4-momenta.
\label{fig:input}}
\end{figure}

Next, we compare the RNN classifier based on a $k_t$ topology on tower
4-momenta with and without image preprocessing. Table~\ref{fig:performance} and
Fig.~\ref{fig:input} show significant gains in not using jet images,
improving ROC AUC from $0.8321$ to $0.8807$ (resp.,
$R_{\epsilon=50\%}$ from $12.7$ to $24.1$) in the case of $k_t$ topologies. In
addition, this result outperforms the MaxOut architecture operating on images by a significant margin.
This suggests that the projection into an image loses information and impacts
classification performance. We suspect the loss of information
to be due to some of the construction steps of jet images
(i.e., pixelisation, rotation, zooming, cropping and normalization).
In particular, all are applied
at the image-level instead of being performed directly on the 4-momenta, which might
induce artefacts due to the lower resolution, particle superposition and aliasing.
By contrast, the RNN is able to work directly with the 4-momenta of a variable-length set of particles,
without any loss of information.
For completeness, we also compare to the performance of a classifier based purely on the single $n$-subjettiness feature $\tau_{21} := \sfrac{\tau_2}{\tau_1}$  and a classifier based on two features (the trimmed mass and $\tau_{21}$)~\cite{Thaler:2010tr}.
In agreement with previous results based on deep learning~\cite{deOliveira:2015xxd,Barnard:2016qma},
we see that our RNN classifier clearly outperforms this variable.

\paragraph{Measurements of the 4-momenta}
The second factor we varied was the source of the 4-momenta.  The \textit{towers} scenario,
corresponds to the case where the 4-momenta come from the  calorimeter
simulation in \texttt{DELPHES}.
 While the calorimeter simulation is simplistic,
the granularity of the towers is quite large ($10^\circ$ in $\phi$) and it
does not take into account that tracking detectors can provide very accurate
momenta measurements for charged particles that can be combined with calorimetry as in the particle flow approach.
Thus, we also consider the \textit{particles}
scenario, which corresponds to an idealized case where the 4-momenta come from
perfectly measured  stable hadrons from \texttt{PYTHIA}.
Table~\ref{fig:performance} and Fig.~\ref{fig:input} show that further gains
could be made with more accurate measurements of the 4-momenta,
improving  e.g. ROC AUC from $0.8807$ to $0.9185$ (resp., $R_{\epsilon=50\%}$ from $24.1$ to $68.3$)
in the case of $k_t$ topologies.
We also considered a case where the 4-momentum came from the \texttt{DELPHES}
particle flow simulation and the data associated with each particle was
augmented with a particle-flow identifier distinguishing $\pm$ charged hadrons,
photons, and neutral hadrons. This is similar in motivation to
Ref.~\cite{Komiske:2016rsd}, but we did not observe any significant gains in
classification performance with respect to the \textit{towers} scenario.

\begin{figure}
\centering
\includegraphics[width=.6\textwidth]{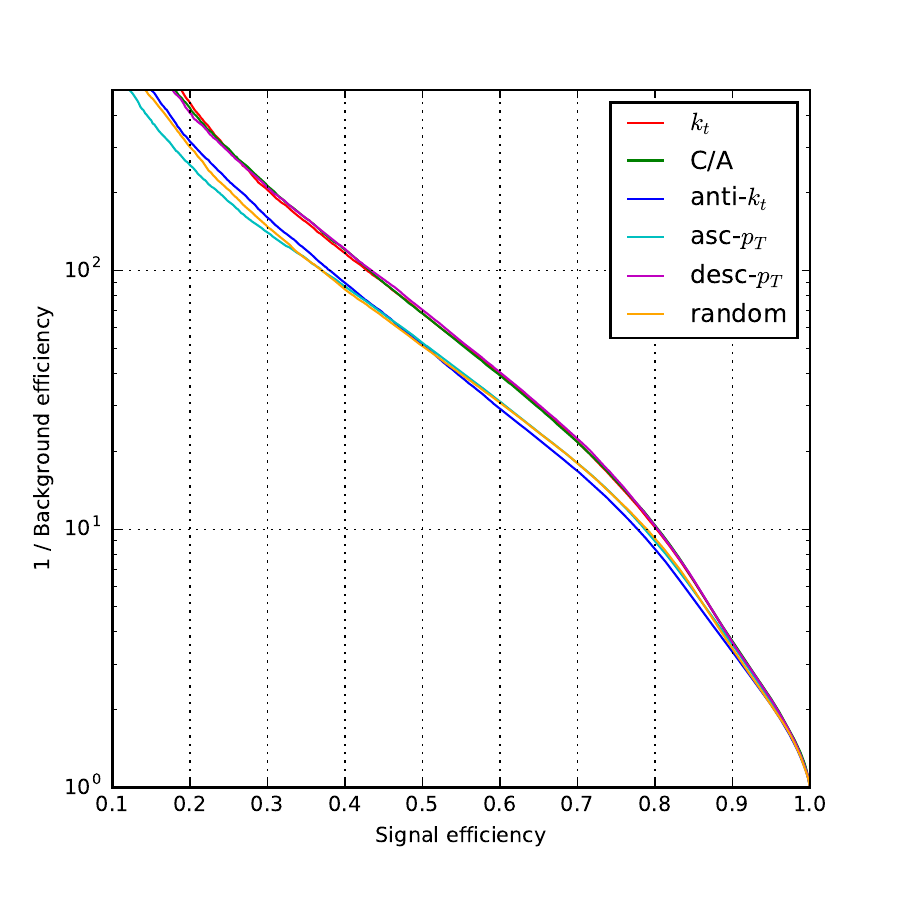}
\caption{Jet classification performance of the RNN classifier based on various network topologies for the embedding (\textit{particles} scenario).
    This plot shows that topology is significant, as supported by the fact that results for $k_t$, C/A and desc-$p_T$
    topologies improve over results for anti-$k_t$, asc-$p_T$ and random binary trees. Best results are achieved for C/A and desc-$p_T$ topologies, depending on the metric considered.
\label{fig:architecture}}
\end{figure}

\paragraph{Topology of the binary trees}
The third factor we studied was the topology of the binary tree $\mathbf{t}_j$
described in Sections~\ref{S:problem} and \ref{sec:re-jets} that dictates the
recursive structure of the RNN.  We considered binary trees based on the anti-$k_t$,
Cambridge-Aachen (C/A), and $k_t$ sequential recombination jet algorithms,
along with random, asc-$p_T$ and desc-$p_T$ binary trees. Table~\ref{fig:performance} and
Fig.~\ref{fig:architecture} show the performance of the RNN classifier based on these various topologies.
Interestingly, the topology is significant.

For instance, $k_t$ and C/A
significantly outperform the anti-$k_t$ topology on both tower and particle
inputs. This is consistent with intuition from previous jet substructure studies
where jets are typically reclustered with the $k_t$ algorithm.  The fact that
the topology is important is further supported by the poor performance of the
random binary tree topology.
% As illustrated in Fig.~\ref{fig:topology}, $k_t$
% (and C/A) lead to fairly balanced trees, while anti-$k_t$ leads to very unbalanced
% trees.
We expected however that a simple sequence (represented as a degenerate binary
tree) based on ascending and descending $p_T$ ordering would not perform
particularly well, particularly since the topology does not use any angular
information. Surprisingly, the simple descending $p_T$ ordering slightly
outperforms the RNNs based on $k_t$ and C/A topologies. The descending $p_T$ network
has the highest $p_T$ 4-momenta near the root of the tree, which we expect to be the most important.
We suspect this is the reason that the descending $p_T$
outperforms the ascending $p_T$ ordering on particles, but this is not supported
by the performance on towers. A similar observation
was already made in the context of natural languages~\cite{bowman2015tree,bowman2016modeling,shi2016does},
where tree-based models have at best only slightly outperformed simpler
sequence-based networks. While recursive networks appear as a principled choice,
it is conjectured that recurrent networks
may in fact be able to discover and implicitly use recursive compositional structure
by themselves, without supervision.

% \begin{figure}
% \includegraphics[scale=0.25]{figures/jet-kt.pdf}
% \includegraphics[scale=0.25]{figures/jet-antikt.pdf}
% \caption{Visualization of the recursive activation over $k_t$ (top) and anti-$k_t$ (bottom) topologies for a same jet. $k_t$ clustering
% usually leads to fairly balanced trees, while anti-$k_t$ clustering often produces binary trees that degenerate to linked lists. Size
% and color indicate the relative importance of a particle (or recombination) in the final decision score of the classifier.
% \label{fig:topology}}
% \end{figure}

\paragraph{Gating}
The last factor that we varied was whether or not to incorporate gating in the
RNN. Adding gating increases the number of parameters to 48,761, but this is
still about 20 times smaller than the number of parameters in the MaxOut
architectures used in previous jet image studies.
Table~\ref{fig:performance} shows the performance of the
various RNN topologies with gating. While results improve significantly
with gating, most notably in terms of $R_{\epsilon=50\%}$, the trends in terms of topologies remain unchanged.

\paragraph{Other variants}
Finally, we also considered a number of other variants. For example, we jointly trained a
classifier with the concatenated embeddings obtained over $k_t$ and anti-$k_t$ topologies, but saw no
significant performance gain. We also tested the performance of recursive
activations transferred across  topologies. For instance, we used the recursive
activation learned with a $k_t$ topology when applied to an anti-$k_t$ topology
and observed a significant loss in performance. We also considered particle and
tower level inputs with an additional trimming preprocessing step, which was
used for the jet image studies, but we saw a significant loss in performance.
While the trimming degraded classification performance, we did not evaluate the
robustness to pileup that motivates trimming and other jet grooming procedures.

\subsection{Infrared and Collinear Safety Studies}

In proposing variables to characterize substructure, physicists have
been equally concerned with classification performance and the ability
to ensure various theoretical properties of those variables. In particular,
initial work on jet algorithms focused on the Infrared-Collinear (IRC) safe conditions:
\begin{itemize}
%    \item \textit{Permutation invariance.} The model should not
%    depend on the order of the particles $\{ \mathbf{v}_j \}$.

    \item \textit{Infrared safety.} The model is robust to augmenting $\mathbf{e}$
    with additional particles $\{ \mathbf{v}_{N + 1}, \dots, \mathbf{v}_{N+K}\}$
    with small transverse momentum.

    \item \textit{Collinear safety.} The model is robust to a collinear
    splitting of a particle, which is represented by replacing a particle $\mathbf{v}_j \in \mathbf{e}$
    with two particles $\mathbf{v}_{j_1}$ and $\mathbf{v}_{j_2}$, such that
    $\mathbf{v}_j = \mathbf{v}_{j_1} + \mathbf{v}_{j_2}$ and
    $\mathbf{v}_{j_1} \cdot \mathbf{v}_{j_2} = ||\mathbf{v}_{j_1}||\,||\mathbf{v}_{j_2}|| - \epsilon$.
\end{itemize}

The sequential recombination algorithms lead to an IRC-safe definition of jets, in the
sense that given the event $\mathbf{e}$, the number of jets $M$ and their 4-momenta $\mathbf{v}(\mathbf{t}_j)$ are IRC-safe.

An early
motivation of this work is that basing the RNN topology on the sequential recombination
algorithms would provide an avenue to machine learning classifiers with some theoretical
guarantee of IRC safety. If one only wants to ensure robustness to only one soft particle or one collinear split,
this could be satisfied by simply running a single iteration of the jet algorithm as a pre-processing step.
However, it is difficult to ensure a more general notion of IRC safety on the embedding due to the non-linearities
in the network. Nevertheless, we can explicitly test the robustness of the embedding or the subsequent classifier to
the addition of soft particles or collinear splits to the input 4-momenta.

Table~\ref{tab:IRC} shows the results of a non-gated RNN trained on the
nominal particle-level input when applied to testing data with additional soft
particles or collinear splits. The collinear splits were uniform in the momentum
fraction and maintained the small invariant mass of the hadrons. We considered
one or ten collinear splits on both random particles and the highest $p_T$
particles. We see that while the 30 models trained with a descending $p_T$
topology very slightly outperform the $k_t$ topology for almost scenarios, their
performance in terms of $R_{\epsilon=50\%}$ decreases relatively more rapidly
when collinear splits are applied (see e.g., the {\it collinear10-max} scenarios
where the performance of $k_t$ decreases by 4\%, while the performance of $p_T$ decreases by 10\%).
This suggests a higher robustness towards collinear
splits for recursive networks based on $k_t$ topologies.

We also point out that the training of these networks is based solely on the classification loss
for the nominal sample. If we are truly concerned with the IRC-safety considerations, then
it is natural to augment the training of the classifiers to be robust to these variations. A number
of modified training procedures exist, including e.g., the adversarial training procedure described in
Ref.~\cite{Louppe:2016ylz}.

\begin{table}
\caption{Performance of pre-trained RNN classifiers (without gating) applied to nominal and modified particle inputs.
The {\it collinear1} ({\it collinear10}) scenarios correspond to applying collinear splits to one (ten) random particles within the jet.
The {\it collinear1-max} ({\it collinear10-max}) scenarios correspond to applying collinear splits to the highest $p_T$ (ten highest $p_T$) particles in the jet.
The {\it soft} scenario corresponds to adding 200 particles with $p_T = 10^{-5}$ GeV uniformly in $0<\phi<2\pi$ and $-5 < \eta < 5$.}
\begin{center}
\begin{tabular}{|c|c|c|c|}
\hline
Scenario & Architecture & ROC AUC & $R_{\epsilon=50\%}$ \\ \hline \hline
                nominal	    & $k_t$ &  0.9185 $\pm$ 0.0006 	& 68.3 $\pm$ 1.8 \\
        nominal 	    & desc-$p_T$ &  0.9189 $\pm$ 0.0009 	& 70.4 $\pm$ 3.6 \\ \hline
            collinear1 	    & $k_t$ &  0.9183 $\pm$ 0.0006 	& 68.7 $\pm$ 2.0 \\
          collinear1 	& desc-$p_T$ &  0.9188 $\pm$ 0.0010 	& 70.7 $\pm$ 4.0 \\ \hline
               collinear10 	& $k_t$ & 0.9174 $\pm$ 0.0006 	& 67.5 $\pm$ 2.6 \\
         collinear10 	& desc-$p_T$ &  0.9178 $\pm$ 0.0011 	& 67.9 $\pm$ 4.3 \\ \hline
           collinear1-max 	& $k_t$ &  0.9184 $\pm$ 0.0006 	& 68.5 $\pm$ 2.8 \\
      collinear1-max 	& desc-$p_T$ &  0.9191 $\pm$ 0.0010 	& 72.4 $\pm$ 4.3 \\ \hline
          collinear10-max 	& $k_t$ &  0.9159 $\pm$ 0.0009 	& 65.7 $\pm$ 2.7 \\
     collinear10-max 	& desc-$p_T$ &  0.9140 $\pm$ 0.0016 	& 63.5 $\pm$ 5.2 \\ \hline
          soft 	            &  $k_t$ & 0.9179 $\pm$ 0.0006 	& 68.2 $\pm$ 2.3 \\
     soft 	            & desc-$p_T$ &  0.9188 $\pm$ 0.0009 	& 70.2 $\pm$ 3.7 \\

\hline
\end{tabular}
\end{center}
\label{tab:IRC}
\end{table}

% ==============================================================================

\section{Experiments with event-level classification}
\label{S:EventResults}

\begin{table}
\caption{Summary of event classification performance. Best results are
achieved through nested recurrence over the jets and over their constituents,
as motivated by QCD.}
\begin{center}
\begin{tabular}{| c |c|c|c|}
\hline
Input  &  ROC AUC & $R_{\epsilon=80\%}$ \\ \hline
\hline    \multicolumn{3}{c}{Hardest jet} \\ \hline
$\mathbf{v}(\mathbf{t}_j)$   	&  0.8909 $\pm$ 0.0007 	& 5.6 $\pm$ 0.0  \\
$\mathbf{v}(\mathbf{t}_j)$, $\mathbf{h}_j^{\text{jet}(k_t)}$  	& \bf 0.9602 $\pm$ 0.0004 	& \bf 26.7 $\pm$ 0.7 \\
$\mathbf{v}(\mathbf{t}_j)$, $\mathbf{h}_j^{\text{jet}(\mathrm{desc}-p_T)}$  	&   0.9594 $\pm$ 0.0010 	& 25.6 $\pm$ 1.4 \\
\hline    \multicolumn{3}{c}{2 hardest jets } \\ \hline
$\mathbf{v}(\mathbf{t}_j)$   	&  0.9606 $\pm$ 0.0011 	& 21.1 $\pm$ 1.1  \\
$\mathbf{v}(\mathbf{t}_j)$, $\mathbf{h}_j^{\text{jet}(k_t)}$  	&  0.9866 $\pm$ 0.0007 	& 156.9 $\pm$ 14.8  \\
$\mathbf{v}(\mathbf{t}_j)$, $\mathbf{h}_j^{\text{jet}(\mathrm{desc}-p_T)}$  	&  \bf 0.9875 $\pm$ 0.0006 	& \bf 174.5 $\pm$ 14.0 \\
 \hline    \multicolumn{3}{c}{5 hardest jets} \\ \hline
$\mathbf{v}(\mathbf{t}_j)$   	&  0.9576 $\pm$ 0.0019 	& 20.3 $\pm$ 0.9  \\
$\mathbf{v}(\mathbf{t}_j)$, $\mathbf{h}_j^{\text{jet}(k_t)}$  	&   0.9867 $\pm$ 0.0004 	& 152.8 $\pm$ 10.4  \\
$\mathbf{v}(\mathbf{t}_j)$, $\mathbf{h}_j^{\text{jet}(\mathrm{desc}-p_T)}$  	& \bf  0.9872 $\pm$ 0.0003 	& \bf 167.8 $\pm$ 9.5\\
 \hline    \multicolumn{3}{c}{No jet clustering, desc-$p_T$ on $\mathbf{v}_i$ } \\ \hline
$i=1$	&  0.6501 $\pm$ 0.0023 	& 1.7 $\pm$ 0.0 \\
$i=1, \dots, 50$	& \bf 0.8925 $\pm$ 0.0079 	& \bf 5.6 $\pm$ 0.5 \\
$i=1, \dots, 100$	&  0.8781 $\pm$ 0.0180 	& 4.9 $\pm$ 0.6  \\
$i=1, \dots, 200$	&  0.8846 $\pm$ 0.0091 	& 5.2 $\pm$ 0.5\\
$i=1, \dots, 400$	&  0.8780 $\pm$ 0.0132 	& 4.9 $\pm$ 0.5  \\
\hline
\end{tabular}
\end{center}
\label{tab:event}
\end{table}

\begin{figure}
\centering
\includegraphics[width=.6\textwidth]{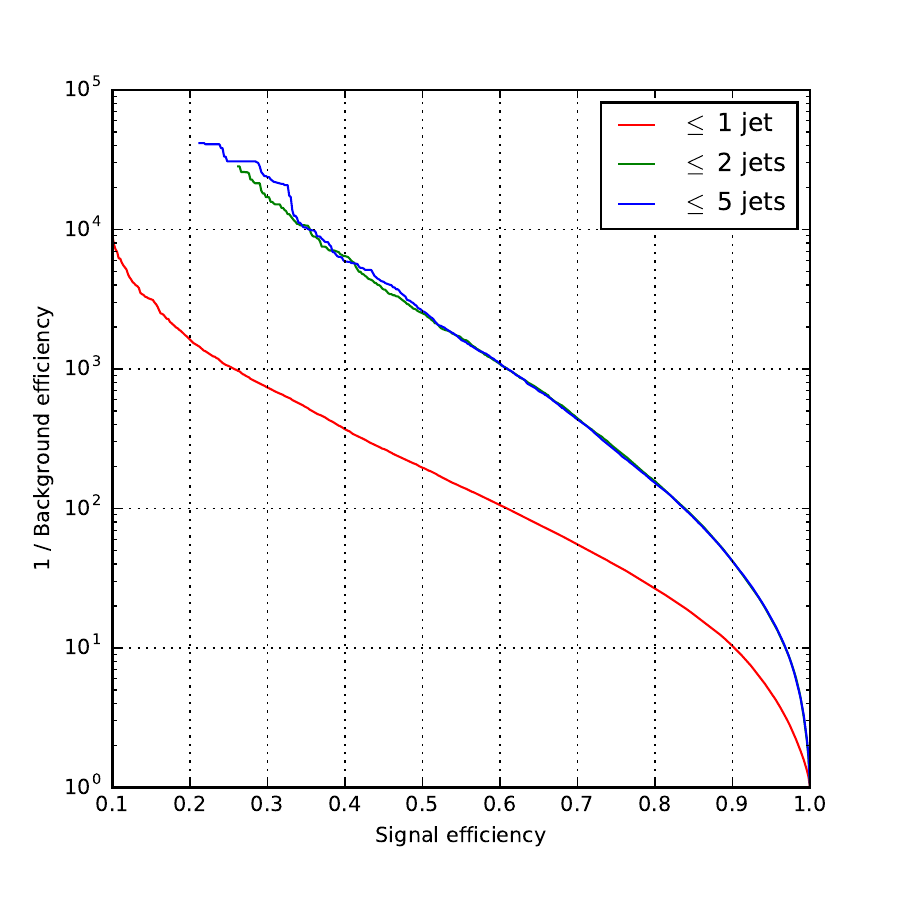}
\caption{
Event classification performance of the RNN classifier when varying
the maximum number of jets given as input to the GRU.
This plots shows there is significant improvement from going to the hardest to the 2 hardest jets, while
there is no to little gain in considering more jets.
\label{fig:event-n-jets}}
\end{figure}

% outline
% describe hyper-parameters
% factor1: vary number of jets in the embeddings
% factor2: kt vs desc-pt
% compare with baselines, emphasize the importance of the nested embedding (vs v(t_j)) and of the grouping by jet (vs v_i)
% mention (omitted) results for towers and towers+pflow

As in the previous section, we carried out a number of performance studies.
However, our goal is mainly to demonstrate the relevance and scalability of the
QCD-motivated approach we propose, rather than making a statement about the
physics reach of the signal process. Results are discussed considering the
idealized \textit{particles} scenario, where the 4-momenta come from perfectly
measured  stable hadrons from \texttt{PYTHIA}. Experiments for the
\textit{towers} scenario (omitted here) reveal similar qualitative conclusions,
though performance was slightly worse for all models, as expected.

\paragraph{Number of jets}
The first factor we varied was the maximum number of jets in the sequence of
embeddings given as input to the GRU. While the event-level embedding can be
computed over all the jets it is constituted by, QCD suggests that the 2 highest
$p_T$ jets hold most of the information to separate signal from background
events, with only marginal discriminating information left in the subsequent
jets. As Table~\ref{tab:event} and Fig.~\ref{fig:event-n-jets} show, there is
indeed significant improvement in going from the hardest jet to the 2 hardest
jets, while there is no to little gain in considering more jets.
Let us also emphasize that the event-level models have only 18,681 parameters,
and were trained on 100,000 training examples.

\paragraph{Topology of the binary trees}
The second factor we studied was the architecture of the networks used for the
inner embedding of the jets, for which we compare $k_t$ against descending $p_T$
topologies. As in the previous section, best results are achieved with descending $p_T$
topologies, though the difference is only marginal.

\paragraph{Other variants}
Finally, we also compare with baselines. With respect to an event-level
embedding computed only from the 4-momenta $\mathbf{v}(\mathbf{t}_j)$ (for $j=1,
\dots, M$) of the jets, we find that augmenting the input to the GRU with
jet-level embeddings yields significant improvement, e.g. improving ROC AUC from $
0.9606 $ to $ 0.9875 $ (resp.  $R_{\epsilon=80\%}$  from $21.1$ to $ 174.5$)
when considering the 2 hardest jets case. This suggests that jet substructures
are important to separate signal from background events, and correctly learned
when nesting embeddings. Similarly, we observe that directly feeding the GRU
with the 4-momenta $\mathbf{v}_i$, for $i=1, \dots, N$, of the constituents of
the event performs significantly worse. While performance remains decent (e.g.,
with a ROC AUC of $0.8925$ when feeding the 50 4-momenta with largest $p_T$),
this suggests that the recurrent network fails to leverage some of the relevant
information, which is otherwise easier to identify and learn when inputs to the
GRU come directly grouped as jets, themselves structured as trees. In contrast
to our previous results for jet-level experiments, this last comparison
underlines the fact that integrating domain knowledge by structuring the
network topology is in some cases crucial for performance.

Overall, this study shows that event embeddings inspired from QCD and produced
by nested recurrence, over the jets and over their constituents, is a promising
avenue for building effective machine learning models. To our knowledge,
this is the first classifier operating at the hadron-level for all
the particles in an event, in a way motivated in its structure by QCD.

% ==============================================================================

\section{Related work}
\label{sec:related}

Neural networks in particle physics have a long history. They have been used in
the past for many tasks, including early work on quark-gluon
discrimination~\cite{Lonnblad:1990bi,Lonnblad:1990qp}, particle
identification~\cite{Sinkus:1996ch}, Higgs tagging~\cite{Chiappetta:1993zv} or
track identification~\cite{Denby:1987rk}. In most of these, neural networks
appear as shallow multi-layer perceptrons where input features were designed by
experts to incorporate domain knowledge. More recently,  the success of deep
convolutional networks has triggered a new body of work in jet
physics,
shifting the paradigm from engineering input features to learning them
automatically from raw data, e.g., as in these works treating jets as images~\cite{Cogan:2014oua,deOliveira:2015xxd,Almeida:2015jua,Baldi:2016fql,Guest:2016iqz,Barnard:2016qma,Komiske:2016rsd,Kasieczka:2017nvn}.
Our work builds instead upon an analogy between QCD and
natural languages, hence complementing the set of algorithms for jet physics
with techniques initially developed for natural language processing~\cite{goller1996learning,socher2011parsing,socher2011semi,cho2014properties,cho2014learning,chen2015sentence}.
In addition, our approach
does not delegate the full modeling task to the machine.
It allows to incorporate domain knowledge in terms of the network architecture,
specifically by structuring the recursion stack for the embedding directly
from QCD-inspired jet algorithms (see Sec.~\ref{sec:re})

Between the time that this work appeared on the arXiv preprint server and submitted for publication there has been a flurry of activity connecting deep learning techniques and jet physics (for reviews see Refs.\cite{Larkoski:2017jix, Guest:2018yhq,Russell:2017cut}). In particular the method described here was also used for quark/gluon tagging in Ref.~\cite{Cheng:2017rdo} and a variant of this method was used to reconstruct a jet's charge~\cite{Fraser:2018ieu}. The authors of Ref.~\cite{Egan:2017ojy} used the tree structure defined by the jet clustering history to define a   substructure ordering scheme for use with a sequential recurrent neural network. Going the opposite direction, graph neural networks and message passing neural networks have also been applied to the the same jet-level classification problem and data described in this work~\cite{Henrion:DLPS2017}.
There has also been a spate of recent work on using QCD-inspired variables, enforcing physical constraints into neural networks, and ensuring infrared safety of neural network based approaches to jet physics~\cite{Butter:2017cot, Datta:2017lxt,Komiske:2017aww,Lim:2018toa,Choi:2018dag}. Exploring a complementary direction, several authors have developed ways to train machine learning techniques using real data to avoid sensitivity to systematic effects in the simulation~\cite{Metodiev:2017vrx, Komiske:2018oaa,Collins:2018epr,DAgnolo:2018cun}. These recent training techniques  are agnostic to the network architecture and can be paired with the RNN architectures described here. Finally, deep learning techniques are now being studied as generative models for jets, where the tree-based model mimics the parton shower and can be trained on real data~\cite{Andreassen:2018apy}. Learning generative modes for both signal and background  classes of jets can be used to define a classifier in which each branch of the tree can be interpreted as a contribution to a likelihood ratio discriminant~\cite{Andreassen:2018apy}.

%\cite{Larkoski:2017jix, Guest:2018yhq,Russell:2017cut}%reviews
%\cite{Butter:2017cot} %Lorentz layer
%\cite{Egan:2017ojy} % LSTM
%\cite{Cheng:2017rdo} %RecNN for q/g
%\cite{Fraser:2018ieu} %compared sequences and RNN
%\cite{Datta:2017lxt,Komiske:2017aww,Lim:2018toa} %physics inspired variables
%\cite{Andreassen:2018apy} %Junipr
%\cite{Nguyen:2018ugw} %topology classification used sequence
%%\cite{Monk:2018zsb} DL as a parton shower
%\cite{Metodiev:2017vrx, Komiske:2018oaa,Collins:2018epr,DAgnolo:2018cun} %learning from data
%\cite{Choi:2018dag} % infrared safety
%%\cite{Luo:2017ncs,Sakaki:2018opq} also q/g... relevant?
%%\cite{DAgnolo:2018cun} new physics searches
%\cite{Henrion:DLPS2017} %MPNN

% ==============================================================================

\section{Conclusions}
\label{sec:conclusions}

Building upon an analogy between QCD and natural languages, we have presented in
this work a novel class of recursive neural networks for jet physics that are
derived from sequential recombination jet algorithms. Our experiments have revealed
that preprocessing steps applied to jet images during their construction (specifically the
pixelisation) loses information,
which impacts classification performance. By contrast, our recursive network is
able to work directly with the four-momenta of a variable-length set of
particles, without the loss of information due to discretization into pixels.
Our experiments indicate that this
results in significant gains in terms of accuracy and data efficiency with
respect to previous image-based networks. Finally, we also showed for the first time
a hierarchical, event-level classification model operating on all the hadrons of an event.
Notably, our results showed that incorporating domain knowledge derived
from jet algorithms and encapsulated in terms of the network architecture led to improved
classification performance.

While we initially expected recursive networks operating on
jet recombination trees to outperform simpler  $p_T$-ordered architectures,
our results still clearly indicate that the topology has an effect on the final
performance of the classifier. However, our initial studies indicate
that architectures based on jet trees are more robust to infrared radiation and collinear splittings than
the simpler  $p_T$-ordered architectures, which may outweigh what at face value appears
to be a small loss in performance.
Accordingly, it would be natural to include robustness to pileup, infrared radiation, and collinear splittings
directly in the training procedure~\cite{Louppe:2016ylz}.
Moreover, it is compelling to think of
generalizations in which the optimization would include the topology used for
the embedding as learnable component instead of considering it fixed a priori.
An immediate challenge of this approach is that a discontinuous change in the
topology (e.g., from varying $\alpha$ or $R$) makes the loss non-differentiable
and rules out standard back propagation optimization algorithms. Nevertheless,
solutions for learning composition orders have recently been proposed in NLP, using
either explicit supervision~\cite{bowman2016fast} or reinforcement
learning~\cite{yogatama2016learning}; both of which could certainly be adapted
to jet embeddings.
Another promising generalization is to use a
graph-convolutional network that operates on a graph where the vertices correspond
to particle 4-momenta $\mathbf{v}_i$ and the edge weights are given by $d_{ii'}^\alpha$ or a similar
QCD-motivated quantity~\cite{DBLP:journals/corr/BrunaZSL13,DBLP:journals/corr/HenaffBL15,
DBLP:journals/corr/LiTBZ15,DBLP:journals/corr/NiepertAK16,DBLP:journals/corr/DefferrardBV16,
kipf2016semi,DBLP:journals/corr/KipfW16, Henrion:DLPS2017}. In conclusion, we feel confident that there is great potential in hybrid techniques like this that incorporate physics knowledge and leverage the power of machine learning.

% ==============================================================================

\begin{acknowledgments}
   We would like to thank the authors of Ref.\citep{Barnard:2016qma} for sharing the data used in their
    studies and Noel Dawe in particular for his responsiveness in clarifying details about their work.
    We would also like to thank Joan Bruna for enlightening discussions about graph-convolutional networks.
    Cranmer and Louppe are both supported through NSF ACI-1450310, additionally Cranmer and Becot are supported
    through PHY-1505463 and PHY-1205376.
\end{acknowledgments}

% ==============================================================================

%\newpage
%\bibliographystyle{hacm}
%\bibliographystyle{hunsrt}
\bibliographystyle{JHEP}
\bibliography{paper.bib}

%\newpage
\appendix

% ==============================================================================

\section{Gated recursive jet embedding}
\label{sec:grs}

The recursive activation proposed in Sec.~\ref{sec:re-jets} suffers from two
critical issues. First, it assumes that left-child, right-child and local node
information $\mathbf{h}^\text{jet}_{k_L}$, $\mathbf{h}^\text{jet}_{k_R}$, $\mathbf{u}_{k}$ are all
equally relevant for computing the new activation, while only some of this
information may be needed and selected. Second, it forces information to pass
through several levels of non-linearities and does not allow to propagate
unchanged from leaves to root. Addressing these issues and generalizing from
\citep{cho2014properties,cho2014learning,chen2015sentence}, we
recursively define a recursive activation equipped with reset and update gates as follows:
\begin{align}
\mathbf{h}^\text{jet}_k &=
 \begin{cases}
  \mathbf{u}_k & \text{if } k \text{ is a leaf} \\
  \mathbf{z}_H \odot \tilde{\mathbf{h}}^\text{jet}_k + \mathbf{z}_L \odot \mathbf{h}^\text{jet}_{k_L} +  & \text{otherwise}\\
  \hookrightarrow \mathbf{z}_R \odot \mathbf{h}^\text{jet}_{k_R}  + \mathbf{z}_N \odot \mathbf{u}_{k} &
 \end{cases} \label{eqn:rec-nn-gated}\\
\mathbf{u}_k &= \sigma \left( W_u g(\mathbf{o}_k) + b_u \right) \\
\mathbf{o}_k &=
\begin{cases}
\mathbf{v}_{i(k)} & \text{if } k \text{ is a leaf} \\
\mathbf{o}_{k_L} + \mathbf{o}_{k_R} & \text{otherwise}
\end{cases}\\
\tilde{\mathbf{h}}^\text{jet}_k &= \sigma \left( W_{\tilde{h}}
\begin{bmatrix}
    \mathbf{r}_L \odot \mathbf{h}^\text{jet}_{k_L} \\
    \mathbf{r}_R \odot \mathbf{h}^\text{jet}_{k_R} \\
    \mathbf{r}_N \odot \mathbf{u}_{k}
\end{bmatrix} + b_{\tilde{h}} \right)\\
\begin{bmatrix}
\mathbf{z}_H \\
\mathbf{z}_L \\
\mathbf{z}_R \\
\mathbf{z}_N
\end{bmatrix} &= \text{softmax} \left( W_z
\begin{bmatrix}
    \tilde{\mathbf{h}}^\text{jet}_k \\
    \mathbf{h}^\text{jet}_{k_L} \\
    \mathbf{h}^\text{jet}_{k_R} \\
    \mathbf{u}_{k}
\end{bmatrix} + b_z
\right) \\
\begin{bmatrix}
\mathbf{r}_L \\
\mathbf{r}_R \\
\mathbf{r}_N
\end{bmatrix} &= \text{sigmoid} \left( W_r
\begin{bmatrix}
    \mathbf{h}^\text{jet}_{k_L} \\
    \mathbf{h}^\text{jet}_{k_R} \\
    \mathbf{u}_{k}
\end{bmatrix} + b_r
\right)
\end{align}
where
$W_{\tilde{h}} \in \mathbb{R}^{q \times 3q}$,
$b_{\tilde{h}} \in \mathbb{R}^q$,
$W_z \in \mathbb{R}^{q \times 4q}$,
$b_z \in \mathbb{R}^q$,
$W_r \in \mathbb{R}^{q \times 3q}$,
$b_r \in \mathbb{R}^q$,
$W_u \in \mathbb{R}^{q \times 4}$
and $b_u \in \mathbb{R}^q$ form together the shared parameters to be learned,
$\sigma$ is the ReLU activation function
and $\odot$ denotes the element-wise multiplication.

Intuitively, the reset gates $\mathbf{r}_L$, $\mathbf{r}_R$ and $\mathbf{r}_N$
control how to actively select and then merge the left-child embedding
$\mathbf{h}^\text{jet}_{k_L}$, the right-child embedding $\mathbf{h}^\text{jet}_{k_R}$ and the local
node information $\mathbf{u}_k$
to form a new candidate activation $\tilde{\mathbf{h}}^\text{jet}_k$.
The final embedding $\mathbf{h}^\text{jet}_k$ can then be regarded as a choice among the candidate
activation, the left-child embedding, the right-child embedding and the local
node information, as controlled by the update gates $\mathbf{z}_H$,
$\mathbf{z}_L$, $\mathbf{z}_R$ and $\mathbf{z}_N$.
Finally, let us note that the proposed gated recursive embedding
is a generalization of Section~\ref{sec:re-jets}, in the sense that the
later corresponds to the case where update gates are set to $\mathbf{z}_H=1$, $\mathbf{z}_L=0$, $\mathbf{z}_R=0$ and $\mathbf{z}_N=0$
and reset gates to $\mathbf{r}_L=1$, $\mathbf{r}_R=1$ and $\mathbf{r}_N=1$ for all nodes $k$.

\section{Gated recurrent event embedding}
\label{sec:gru}

In this section, we  formally define the gated recurrent event embedding introduced in Sec.~\ref{sec:re-events}.
Our event embedding function is a GRU
\citep{chung2014empirical} operating on the $p_T$ ordered sequence of pairs
$(\mathbf{v}(\mathbf{t}_j), \mathbf{h}^\text{jet}_1(\mathbf{t}_j))$, for $j=1, \dots, M$,
where $\mathbf{v}(\mathbf{t}_j)$ is the unprocessed 4-momentum $(\phi, \eta, p_T, m)$ of the jet $\mathbf{t}_j$
and $\mathbf{h}^\text{jet}_1(\mathbf{t}_j)$ is its embedding.
Its final output $\mathbf{h}^\text{event}_{j=M}$ is recursively defined as follows:
\begin{align}
\mathbf{h}^\text{event}_j &= \mathbf{z}_j \odot \mathbf{h}^\text{event}_{j-1} + (1 - \mathbf{z}_j) \odot \tilde{\mathbf{h}}^\text{event}_j\\
\tilde{\mathbf{h}}^\text{event}_j &= \sigma \left( W_{hx} \mathbf{x}_j + W_{hh} (\mathbf{r}_j \odot \mathbf{h}^\text{event}_{j-1}) + b_h \right)\\
\mathbf{x}_j &= \begin{bmatrix}
    \mathbf{v}(\mathbf{t}_j) \\
    \mathbf{h}^\text{jet}_1(\mathbf{t}_j)
\end{bmatrix} \\
\mathbf{z}_j &= \text{sigmoid} \left(  W_{zx} \mathbf{x}_j + W_{zh}  \mathbf{h}^\text{event}_{j-1} + b_z \right) \\
\mathbf{r}_j &= \text{sigmoid} \left(  W_{rx} \mathbf{x}_j + W_{rh}  \mathbf{h}^\text{event}_{j-1} + b_r \right)
\end{align}
where
$W_{hx} \in \mathbb{R}^{r \times 4+q}$,
$W_{hh} \in \mathbb{R}^{r \times r}$,
$b_h \in \mathbb{R}^r$,
$W_{rx} \in \mathbb{R}^{r \times 4+q}$,
$W_{rh} \in \mathbb{R}^{r \times r}$,
$b_r \in \mathbb{R}^r$,
$W_{zx} \in \mathbb{R}^{r \times 4+q}$,
$W_{zh} \mathbb{R}^{r \times r}$ and
$b_z \in \mathbb{R}^r$
are the parameters of the embedding function,
$r$ is the size of the embedding,
$\sigma$ is the ReLU activation function,
and $\mathbf{h}^\text{event}_0 = 0$.
In the experiments of Sec.~\ref{S:EventResults}, only the 1, 2 or 5 hardest
jets are considered in the sequence $j=1, \dots, M$, as ordered by ascending values of $p_T$.

\section{Implementation details}
\label{sec:impl}

While tree-structured networks appear to be a principled choice in natural
language processing, they often have been overlooked in favor of sequence-based
networks on the account of their technical incompatibility with batch
computation~\cite{bowman2016fast}. Because tree-structured networks use a
different topology for each example,  batching is indeed often impossible in
standard implementations, which prevents them from being trained efficiently on
large datasets. For this reason, the recursive jet embedding we introduced in
Sec.~\ref{sec:re-jets} would undergo the same technical issues if not
implemented with caution.

In our implementation, we achieve batch computation by noticing that activations
from a same level in a recursive binary tree can be performed all at once,
provided all necessary computations from the deeper levels have already been
performed. This principle extends to the synchronized computation across
multiple trees, which enables the batch computation of our jet embeddings
across many events. More specifically, the computation of jet embeddings is
preceded by a traversal of the recursion trees for all jets in the batch, and whose
purpose is to unroll and regroup computations by their level of recursion. Embeddings are
then reconstructed level-wise in batch, in a bottom-up fashion, starting from the deepest level
of recursion across all trees.

Finally, learning is carried out through gradients obtained by the automatic
differentiation of the full model chain on a batch of events (i.e., the recursive computation of jet embeddings,
the sequence-based recurrence to form the event embeddings, and the forward pass through the classifier).
The implementation is written in native Python code and makes use of Autograd~\cite{autograd}
for the easy derivation of the gradients over dynamic structures.
Code is available at \footnote{\url{https://github.com/glouppe/recnn}}
under BSD license for further technical details.

% ==============================================================================

\end{document}